\documentclass[12pt]{article}
\begin{document}
\newcommand{\beq}{\begin{equation}}
\newcommand{\eeq}{\end{equation}}
\newcommand{\beqa}{\begin{eqnarray}}
\newcommand{\eeqa}{\end{eqnarray}}
\newcommand{\beqar}{\begin{eqnarray*}}
\newcommand{\eeqar}{\end{eqnarray*}}
\newcommand{\al}{\alpha}
\newcommand{\be}{\beta}
\newcommand{\del}{\delta}
\newcommand{\D}{\Delta}
\newcommand{\eps}{\epsilon}
\newcommand{\ga}{\gamma}
\newcommand{\Ga}{\Gamma}
\newcommand{\ka}{\kappa}
\newcommand{\nn}{\nonumber}
\newcommand{\inn}{\!\cdot\!}
\newcommand{\h}{\eta}
\newcommand{\ii}{\iota}
\newcommand{\kk}{\varphi}
\newcommand\F{{}_3F_2}
\newcommand{\la}{\lambda}
\newcommand{\La}{\Lambda}
\newcommand{\na}{\prt}
\newcommand{\Om}{\Omega}
\newcommand{\om}{\omega}
\newcommand{\p}{\phi}
\newcommand{\sig}{\sigma}
\renewcommand{\t}{\theta}
\newcommand{\z}{\zeta}
\newcommand{\ssc}{\scriptscriptstyle}
\newcommand{\eg}{{\it e.g.,}\ }
\newcommand{\ie}{{\it i.e.,}\ }
\newcommand{\labell}[1]{\label{#1}} 
\newcommand{\reef}[1]{(\ref{#1})}
\newcommand\prt{\partial}
\newcommand\veps{\varepsilon}
\newcommand{\pol}{\varepsilon}
\newcommand\vp{\varphi}
\newcommand\ls{\ell_s}
\newcommand\dS{\dot{\cal S}}
\newcommand\dB{\dot{B}}
\newcommand\dG{\dot{G}}
\newcommand\ddG{\dot{\dot{G}}}
\newcommand\dP{\dot{\Phi}}
\newcommand\cF{{\cal F}}
\newcommand\cA{{\cal A}}
\newcommand\cS{{\cal S}}
\newcommand\cT{{\cal T}}
\newcommand\cV{{\cal V}}
\newcommand\cL{{\cal L}}
\newcommand\cM{{\cal M}}
\newcommand\cN{{\cal N}}
\newcommand\cG{{\cal G}}
\newcommand\cH{{\cal H}}
\newcommand\cI{{\cal I}}
\newcommand\cJ{{\cal J}}
\newcommand\cl{{\iota}}
\newcommand\cP{{\cal P}}
\newcommand\cQ{{\cal Q}}
\newcommand\cg{{\it g}}
\newcommand\cR{{\cal R}}
\newcommand\cB{{\cal B}}
\newcommand\cO{{\cal O}}
\newcommand\tcO{{\tilde {{\cal O}}}}
\newcommand\bg{\bar{g}}
\newcommand\bb{\bar{b}}
\newcommand\bH{\bar{H}}
\newcommand\bX{\bar{X}}
\newcommand\bK{\bar{K}}
\newcommand\bA{\bar{A}}
\newcommand\bZ{\bar{Z}}
\newcommand\bxi{\bar{\xi}}
\newcommand\bphi{\bar{\phi}}
\newcommand\bpsi{\bar{\psi}}
\newcommand\bprt{\bar{\prt}}
\newcommand\bet{\bar{\eta}}
\newcommand\btau{\bar{\tau}}
\newcommand\bnabla{\bar{\nabla}}
\newcommand\hF{\hat{F}}
\newcommand\hA{\hat{A}}
\newcommand\hT{\hat{T}}
\newcommand\htau{\hat{\tau}}
\newcommand\hD{\hat{D}}
\newcommand\hf{\hat{f}}
\newcommand\hg{\hat{g}}
\newcommand\hp{\hat{\phi}}
\newcommand\hi{\hat{i}}
\newcommand\ha{\hat{a}}
\newcommand\hb{\hat{b}}
\newcommand\hQ{\hat{Q}}
\newcommand\hP{\hat{\Phi}}
\newcommand\hS{\hat{S}}
\newcommand\hX{\hat{X}}
\newcommand\tL{\tilde{\cal L}}
\newcommand\hL{\hat{\cal L}}
\newcommand\tG{{\widetilde G}}
\newcommand\tg{{\widetilde g}}
\newcommand\tphi{{\widetilde \phi}}
\newcommand\tPhi{{\widetilde \Phi}}
\newcommand\td{{\tilde d}}
\newcommand\tk{{\tilde k}}
\newcommand\tf{{\tilde f}}
\newcommand\ta{{\tilde a}}
\newcommand\tb{{\tilde b}}
\newcommand\tc{{\tilde c}}
\newcommand\tR{{\tilde R}}
\newcommand\teta{{\tilde \eta}}
\newcommand\tF{{\widetilde F}}
\newcommand\tK{{\widetilde K}}
\newcommand\tE{{\widetilde E}}
\newcommand\tpsi{{\tilde \psi}}
\newcommand\tX{{\widetilde X}}
\newcommand\tD{{\widetilde D}}
\newcommand\tO{{\widetilde O}}
\newcommand\tS{{\tilde S}}
\newcommand\tB{{\widetilde B}}
\newcommand\tA{{\widetilde A}}
\newcommand\tT{{\widetilde T}}
\newcommand\tC{{\widetilde C}}
\newcommand\tV{{\widetilde V}}
\newcommand\thF{{\widetilde {\hat {F}}}}
\newcommand\Tr{{\rm Tr}}
\newcommand\tr{{\rm tr}}
\newcommand\STr{{\rm STr}}
\newcommand\hR{\hat{R}}
\newcommand\M[2]{M^{#1}{}_{#2}}

\newcommand\bS{\textbf{ S}}
\newcommand\bI{\textbf{ I}}
\newcommand\bJ{\textbf{ J}}

\begin{titlepage}
\begin{center}

\vskip 2 cm
{\LARGE \bf  $O(9,9)$ symmetry of NS-NS couplings  \\ \vskip 0.75  cm   at order $\alpha'^3$     }\\
\vskip 1.25 cm
   Mohammad R. Garousi\footnote{garousi@um.ac.ir}

\vskip 1 cm
{{\it Department of Physics, Faculty of Science, Ferdowsi University of Mashhad\\}{\it P.O. Box 1436, Mashhad, Iran}\\}
\vskip .1 cm
 \end{center}

\begin{abstract}
Recently, imposing the $O(1,1)$ symmetry on  the circle reduction of  the  classical  effective action  of string theory, we have found all NS-NS couplings  of type II superstring theories at order $\alpha'^3$.  In this paper we use the cosmological reduction on the couplings and show that, up to one-dimensional field redefinitions and total derivative terms, they are invariant under the $O(9,9)$ transformations. 
\end{abstract}
\end{titlepage}

\section{Introduction}

A  theory of gravity in a spacetime manifold with/without boundary which is consistent with the rules of quantum mechanics is the string theory. This theory includes  the finite number of massless modes and the tower of infinite number of  massive modes of the string excitations. At  low energies, however,  the massive modes are integrated out to produce an effective theory which includes only the massless fields.  The effective action  has a double expansions. The genus-expansion which includes  the  classical tree-level  \ie $\bS_{\rm eff}+\prt\!\! \bS_{\rm eff}$  and a tower of  quantum  loop-level  corrections, and the   stringy-expansion which is an expansion in terms of  higher derivative couplings  at each loop level.  The classical    effective action has the following  higher-derivative or  $\alpha'$-expansion:
\beqa
\bS_{\rm eff}&=&\sum^\infty_{m=0}\alpha'^m\bS_m=\bS_0+\alpha' \bS_1 +\alpha'^2 \bS_2+\alpha'^3 \bS_3+\cdots \labell{seff}\\
\prt\!\!\bS_{\rm eff}&=&\sum^\infty_{m=0}\alpha'^m\prt\!\!\bS_m=\prt\!\!\bS_0+\alpha' \prt\!\!\bS_1+\alpha'^2 \prt\!\!\bS_2+\alpha'^3 \prt\!\!\bS_3 +\cdots \nn
\eeqa 
The leading order bulk action $\!\!\bS_0$   includes the Hilbert-Einstein term and   the boundary action $\prt\!\!\bS_0$  includes the Hawking-Gibbons term \cite{York:1972sj,Gibbons:1976ue}. These actions and their appropriate  higher derivative extensions should be found by specific techniques in the string theory. Since the effective action includes the couplings at all orders of derivative, one complication in finding the couplings in the string theory is the freedom of  the field redefinitions that include the higher derivatives of fields \cite{Metsaev:1987zx}. As a result, the effective action in the string theory can appear in many different equivalent schemes.

One of the most exciting discoveries in perturbative  string theory is   T-duality \cite{Giveon:1994fu,Alvarez:1994dn}  which appears  when one compactifies theory  on a torus, \eg the compactification of the full bosonic string theory on tours $T^d$ is invariant under   $O(d,d,Z)$ transformations. After integrating out the massive modes, however, the T-duality  should appears as symmetry in the  effective actions. It has been shown in \cite{Sen:1991zi,Hohm:2014sxa} that the dimensional reduction of the classical effective actions of the bosonic and heterotic string theories on a torus $T^d$ are in fact invariant under  $O(d,d,R)$ transformations. 

When one reduces the effective action  on a circle, the invariance of the reduced action under the $Z_2$-subgroup of the $O(1,1,R)$-group constrains greatly the  couplings in the  effective action. In fact  there is only one T-dual multiplet in the effective action of the bosonic  string theory, and there are two T-dual multiplets in the effective action of type II superstring theories  at the leading order of $\alpha'$, one for NS-NS couplings and one for R-R couplings \cite{Garousi:2019jbq}. By the T-dual multiplet we means the  set of couplings in the effective action which are related into each others under the $Z_2$-transformations after  reducing them  on the circle. The $Z_2$-transformations or T-duality transformations are the Buscher rules \cite{ Buscher:1987sk,Buscher:1987qj} and some  higher derivative corrections at each order of $\alpha'$ which depend on the scheme that one uses for the gauge invariant  couplings in the effective action at that  order of $\alpha'$ \cite{Kaloper:1997ux,Garousi:2019wgz}. The corrected transformations, however,  should satisfy the $Z_2$-symmetry. There is no scheme for the higher-derivative couplings in the original action in which the T-duality transformations are the standard Buscher rules \cite{Garousi:2019wgz}.

Since the  T-duality transformations have higher derivative corrections, the other T-dual multiplets  include couplings at all orders of $\alpha'$. In fact it has been observed in \cite{Garousi:2019mca} that the couplings in the effective action of the bosonic string theory at order $\alpha'$ are related by the T-duality transformations to the couplings at order  $\alpha'^2$. They belong to one T-dual multiplet. If one extends the calculations in \cite{Garousi:2019mca} to the order $\alpha'^3$, one would find the couplings at order $\alpha'$ and $\alpha'^2$ are related to  some of the couplings at order $\alpha'^3$. They belong to the same  T-dual multiplet. However, there are couplings at this order that are not connected to the couplings at order $\alpha',\alpha'^2$ by the T-duality transformations. They belong to another T-dual multiplet.  In other words, if one finds the couplings in the bosonic string theory at order $\alpha'^3$ by S-matrix method, one would find they  have two factors $a_1,a_2$, \ie $\!\bS_3=a_1\!\bS_3^1+a_2\!\bS_3^2$. One factor should be the same as the one appears in the couplings at order $\alpha',\alpha'^2$, and another one which is proportional to  $\z(3)$. These two factors should appear in the T-duality transformations as well. At the order $\alpha'^4$, the couplings in the bosonic theory should have three factors $a_1,a_2,a_3$,  \ie $\!\bS_4=a_1\!\bS_4^1+a_2\!\bS_4^2+a_3\!\bS_4^3$. One is the same as the factor in $\alpha',\alpha'^2$, one which is proportional to $\z(3)$ and another independent factor. Similar pattern  should appear for the higher orders of $\alpha'$. Schematically, the bulk action \reef{seff} has the following expansion in terms of the  T-dual multiplets:
\beqa
\bS_{\rm eff}&=&\sum^\infty_{n=0}a_n T_n=a_0T_0+a_1 T_1+a_2T_2+a_3T_3+\cdots \labell{Teff}
\eeqa
 where $a_0,a_1,\cdots$ are some coefficients that can not be fixed by the T-duality. They should be fixed by the S-matrix calculations, \eg $a_0=1$, $a_1=1$, $a_3=\z(3)$. In the type II superstring theory, $a_1=0$ and there are T-dual multiplets which include the R-R couplings. The T-dual mutiplets in the bosonic string theory are
 \beqa
 T_0&=&\bS_0\nn\\
 T_1&=&\alpha' \bS_1+\alpha'^2\bS_2+\alpha'^3\bS_3^1+\alpha'^4\bS_4^1+\alpha'^5\bS_5^1+\cdots\nn\\
  T_2&=&\alpha'^3 \bS_3^2+\alpha'^4\bS_4^2+\alpha'^5\bS_5^2+\alpha'^6\bS_6^2+\alpha'^7\bS_7^2+\cdots\nn\\
  T_3&=&\alpha'^4 \bS_4^3+\alpha'^5\bS_5^3+\alpha'^6\bS_6^3+\alpha'^7\bS_7^3+\alpha'^8\bS_8^3+\cdots\\
  \vdots\nn
 \eeqa
There are no parameters in these multiplets. If the spacetime manifold has no boundary in which the total derivative terms  can be ignored, then each  multiplet should be  invariant under the T-duality transformations after reducing it on the circle. In principle, this constraint may fix all couplings in each T-dual multiplet. The couplings in the multiplet $T_1$ at orders $\alpha'$ and $\alpha'^2$ in a particular scheme have been found in \cite{Garousi:2019wgz,Garousi:2019mca}.   The couplings in the multiplet $T_2$ at order $\alpha'^3$  have been also found in \cite{Garousi:2020gio,Garousi:2020lof}.

When the spacetime has boundary, however, one should keep the total derivative terms before and after reduction and use  the Stokes's theorem to transfer them  to the boundary. They dictate that the invariance under the T-duality transformations requires some couplings on the boundary  as well \cite{Garousi:2019xlf}.   Hence, the bulk T-dual multiplets should be accompanied  with appropriate boundary couplings to be fully invariant under the T-duality. Schematically, the boundary action \reef{seff} should have  the following expansion in terms of the boundary T-dual multiplets:
\beqa
\prt\!\!\bS_{\rm eff}&=&\sum^\infty_{n=0}a_n \prt T_n=a_0\prt T_0+a_1 \prt T_1+a_2 \prt T_2+a_3\prt T_3+\cdots \labell{Teffb}
\eeqa
 where $a_0,a_1,\cdots$ are the same coefficients that appear in the bulk T-dual multiplets \reef{Teff}.  The boundary  mutiplets are
 \beqa
 \prt T_0&=&\prt\!\!\bS_0\nn\\
\prt  T_1&=&\alpha' \prt\!\!\bS_1+\alpha'^2\prt\!\!\bS_2+\alpha'^3\prt\!\!\bS_3^1+\alpha'^4\prt\!\!\bS_4^1+\alpha'^5\prt\!\!\bS_5^1+\cdots\nn\\
 \prt T_2&=&\alpha'^3 \prt\!\!\bS_3^2+\alpha'^4\prt\!\!\bS_4^2+\alpha'^5\prt\!\!\bS_5^2+\alpha'^6\prt\!\!\bS_6^2+\alpha'^7\prt\!\!\bS_7^2+\cdots\nn\\
 \prt T_3&=&\alpha'^4 \prt\!\!\bS_4^3+\alpha'^5\prt\!\!\bS_5^3+\alpha'^6\prt\!\!\bS_6^3+\alpha'^7\prt\!\!\bS_7^3+\alpha'^8\prt\!\!\bS_8^3+\cdots\\
  \vdots\nn
 \eeqa
The combination of bulk and boundary multiplets, \ie $ T_i+\prt T_i$, are then invariant under the T-duality transformations. In other words,  neither the bulk multiplets nor the boundary multiplets are invariant separately under the T-duality transformations. Their anomalies cancel each other.  There are, however,  boundary couplings that are invariant under the T-duality transformations without anomaly. Some of them are  related to the anomalous boundary multiplets by  
imposing the principle of the least action in the presence of the boundary with appropriate boundary values for the massless fields.  Using these constraints, the boundary coupling in the multiplet $\prt T_0$ has been found in \cite{Garousi:2019xlf,Garousi:2021cfc}.   The boundary couplings in the multiplet $\prt T_1$ at order $\alpha'$ in a particular scheme  have been found in \cite{Garousi:2021cfc}. 

When one uses the cosmological reduction on the classical effective action, the resulting one-dimensional effective action should have $O(d,d,R)$ symmetry \cite{Sen:1991zi,Hohm:2014sxa}. This symmetry has been first observed for the leading order bulk couplings in \cite{Veneziano:1991ek,Meissner:1991zj,Maharana:1992my}   and  for the couplings at order $\alpha'$ in a specific scheme in  \cite{Meissner:1996sa}. The T-duality transformations or $O(d,d,R)$ transformations  in this case also recive higher derivative corrections. The corrected transformations satisfy the $O(d,d,R)$ symmetry \cite{Meissner:1996sa}. In this case also there is no scheme for the original couplings in which the T-duality transformation are the standared $O(d,d,R)$-transformations of the leading order.  Unlike the circle reduction,   some of the couplings in the original action   disappear upon the  reduction \cite{Hohm:2015doa}.  Hence, this symmetry is not appropriate for fixing the couplings in the original action. However, this symmetry is useful for classifying the couplings in the  one-dimensional effective action at all orders of $\alpha'$ \cite{Hohm:2015doa,Hohm:2019jgu}. 

Using the most general corrections for the  T-duality transformations, including the transformations for the lapse function, and  using  integration by part, it has been shown in \cite{Hohm:2015doa,Hohm:2019jgu} that the cosmological reduction of the bulk action \reef{seff} at order $\alpha'$ and higher, can be written in a scheme in which only  the first time-derivative of the generalized metric ${\cal S}$ appears. Trace of odd number of the first-derivative of $\cS$  is zero. It has been shown in \cite{Hohm:2019jgu} that the couplings which include  $\tr(\dS^2)$ can be removed by the lapse function transformation. Then the one-dimensional bulk action can be written in a specific scheme as  the following expansion \cite{Hohm:2015doa,Hohm:2019jgu}:
\beqa
\bS_{\rm eff}^c&=&\bS_0^c+\int dt e^{-\Phi}\bigg(\alpha' c_{2,0}\tr(\dS^4)+\alpha'^2c_{3,0}\tr(\dS^6)\nn\\
&&\qquad\quad\qquad\quad+\alpha'^3[c_{4,0}\tr(\dS^8)+c_{4,1}(\tr(\dS^4))^2]\nn\\
&&\qquad\quad\qquad\quad+\alpha'^4[c_{5,0}\tr(\dS^{10})+c_{5,1}\tr(\dS^6)\tr(\dS^4)]+\cdots\bigg)\labell{cosm}
\eeqa
where the coefficients $c_{m,n}$ depends on the specific theory, \eg $c_{2,0}$ is non-zero for the bosonic string theory whereas this number is zero for the superstring theory.

To find the cosmological reduction of the corresponding boundary action in \reef{seff}, one has to take into account the one-dimensional total derivative terms and the T-duality transformations or the field redefinitions that have been used in \cite{Hohm:2015doa,Hohm:2019jgu}. However, it has been observed in \cite{Garousi:2021cfc} that if one adds the total derivative term resulting from the cosmological reduction of the leading order action to the boundary by using the Stokes's theorem, it cancels the cosmological reduction of the Hawking-Gibbons term, \ie $\!\bS_0^c=0$.  Since the cosmological reduction of the boundary term at the leading order of $\alpha'$ is zero, we expect it should be zero at all higher orders of $\alpha'$ as well, \ie
\beqa
\prt\!\!\bS_{\rm eff}^c&=&0
\eeqa
This may be used for conforming the boundary couplings in the effective action \reef{seff} at each order of $\alpha'$ that can be found by the $Z_2$-symmetry. Since the cosmological reduction of the leading order action is zero, then for studying the cosmological reduction of the boundary action at order $\alpha'$, one does not need the field redefinitions  used in the bulk action. It has been shown in \cite{Garousi:2021cfc} that if one adds the total derivative terms at order $\alpha'$ to the reduction of the boundary couplings at order $\alpha'$,  they  become zero which is consistent with the above conjecture. 

The cosmological bulk action \reef{cosm} may also be used for confirming the bulk couplings in the effective action \reef{seff} that can be found by the $Z_2$-symmetry. The NS-NS couplings in the effective action of type II superstring theory at order $\alpha'^3$ have been found in \cite{Garousi:2020gio,Garousi:2020lof} by imposing the $Z_2$-symmetry on the effective action. In fact it has been shown in \cite{Garousi:2020mqn} that there are 872 independent couplings at this order. The  $Z_2$-symmetry is imposed on these couplings in a particular scheme. Interestingly, all parameters are fixed up to an overall factor in \cite{Garousi:2020gio}. In that scheme there are 445 non-zero couplings which include derivatives of dilaton. A field redefinition has been used in \cite{Garousi:2020lof} to write them in terms of 251 couplings in which the dilaton appears only as the overall factor $e^{-2\phi}$. In this paper we are going to show that the cosmological reduction of the couplings in \cite{Garousi:2020gio} or in \cite{Garousi:2020lof} can be written in the form of the  cosmological action \reef{cosm} at order $\alpha'^3$. The gravitational sector of these couplings which have been found a long time ago by the S-matrix and sigma model calculations \cite{ Gross:1986iv,Gross:1986mw,Grisaru:1986vi,Freeman:1986zh}, has been shown in \cite{Codina:2020kvj} that satisfies the $O(9,9)$ symmetry.

The outline of the paper is as follows:  In section 2, we review the observation that the cosmological reduction of the  leading order  bulk and boundary actions are invariant under the $O(d,d)$ transformations.   In section 3, we first reduce the bulk NS-NS couplings at order $\alpha'^3$ that have been found in \cite{Garousi:2020lof} to find its corresponding  one-dimensional bulk action. We  add to it all one-dimensional total derivative terms and all possible field redefinitions with arbitrary coefficients  to write the action in a scheme   which has all the arbitrary parameters of the field redefinitions and the total derivative terms. We then  impose the constraint on the parameters that the cosmological action has no derivative of the one-dimensional dilaton, no second and higher derivatives on metric and $B$-field and we impose the constraint that the couplings involving the first derivative of metric which are not consistent with the $O(9,9)$ symmetry to be zero. Moreover, we impose the condition that the terms which have contribution to the $O(9,9)$-invariant structure $\tr(\dS^2)$ to be zero. We have found that  in fact there is a solution for the parameters for such conditions. After imposing the resulting relations between the parameters, we find the  action in the scheme which can be written explicitly as the $\alpha'^3$-order terms of  \reef{cosm}. 

\section{Cosmological reduction at the leading order}

In this section, we review the cosmological reduction of the leading order bulk and boundary actions. These actions are given as 
\beqa
\bS_0+\prt\!\!\bS_0
=-\frac{2}{\kappa^2}\Bigg[ \int d^{D}x \sqrt{-G} e^{-2\phi} \left(  R + 4\nabla_{\mu}\phi \nabla^{\mu}\phi-\frac{1}{12}H^2\right)+ 2\int d^{D-1}\sigma\sqrt{| g|}  e^{-2\phi}K\Bigg]\labell{baction}
\eeqa
where $G$ is determinant of the bulk metric $G_{\mu\nu}$ and boundary is specified by the functions $x^\mu=x^\mu(\sigma^{\tilde{\mu}})$. In the boundary action, $g$ is determinant of the induced metric on the boundary 
\beqa
g_{\tilde{\mu}\tilde{\nu}}&=& \frac{\prt x^\mu}{\prt \sigma^{\tilde{\mu}}}\frac{\prt x^\nu}{\prt \sigma^{\tilde{\nu}}}G_{\mu\nu}\labell{indg}
\eeqa
 The extrinsic curvature  of boundary, \ie  $K_{\mu\nu}$, is defined as $
K_{\mu\nu}=\nabla_{\mu}n_{\nu}-n_{\mu}a_\mu$
where $n^\mu$ is the unite vector orthogonal to the boundary which is outward-going (inward-going) if the boundary is spacelike (timelike), and $a_\nu= n^{\rho}\nabla_{\rho}n_{\nu}$ is acceleration. It satisfies the relation $n^\mu a_\mu=0$. The extrinsic curvature is symmetric and satisfies $n^\mu K_{\mu\nu}=0$.

When fields depend only on time, using the gauge symmetries it is possible to write the metric, $B$-field  and dilaton as
 \beqa
G_{\mu\nu}=\left(\matrix{-n^2(t)& 0&\cr 0&G_{ij}(t)&}\right),\, B_{\mu\nu}= \left(\matrix{0&0\cr0&B_{ij}(t)&}\right),\,  2\phi=\Phi+\frac{1}{2}\log\det(G_{ij})\labell{creduce}\eeqa
where the lapse function $n(t)$ can also be fixed to $n=1$. The cosmological reduction of the bulk action then becomes
\beqa
\bS_0^c&=&-\frac{2}{\kappa}\int dt e^{-\Phi}\Bigg[\frac{1}{4}\dB_{ij}\dB^{ij}-\frac{3}{4}\dG_{ij}\dG^{ij}-G^{ij}\dG_{ij}\dP-\dP^2+G^{ij}\ddot{G}_{ij}\Bigg]
\eeqa
where $\dG^{ij}\equiv G^{ik}G^{il}\dG_{kl}$. Using the following total derivative term:
\beqa
\int dt \frac{d}{dt}\Bigg[e^{-\Phi}G^{ij}\dG_{ij}\Bigg]=\int dt e^{-\Phi}\Bigg[-G^{ij}\dG_{ij}\dP-\dG^{ij}\dG_{ij}+G^{ij}\ddot{G}_{ij}\Bigg]
\eeqa
one can write $\bS_0^c$ as
\beqa
\bS_0^c&=&-\frac{2}{\kappa^2}\int dt e^{-\Phi}\Bigg[\frac{1}{4}\dB_{ij}\dB^{ij}+\frac{1}{4}\dG_{ij}\dG^{ij}-\dP^2\Bigg]-\frac{2}{\kappa}\int dt \frac{d}{dt}\Bigg[e^{-\Phi}G^{ij}\dG_{ij}\Bigg]
\eeqa
The total derivative term can be transferred to the boundary by using the Stokes's theorem.

On the other hand, the cosmological boundary is specified by $x^i=\sigma^i$, and $x^0=t$ is independent of $\sigma^i$. Hence, $\sqrt{| g|}  e^{-2\phi}=e^{-\Phi}$. The unit vector to the boundary is fixed, \ie $\dot{n}=0$, and the reduction of the trace of the  extrinsic curvature becomes 
\beqa
K^c=\frac{1}{2}G^{ij}\dG_{ij}\labell{cK}
\eeqa
Therefore, the reduction of the boundary term is cancelled with the total derivative term in the bulk action, \ie
\beqa
\bS_0^c&=&-\frac{2}{\kappa^2}\int dt e^{-\Phi}\Bigg[\frac{1}{4}\dB_{ij}\dB^{ij}+\frac{1}{4}\dG_{ij}\dG^{ij}-\dP^2\Bigg]\labell{S0c}\\
\prt\!\!\bS_0^c&=&0\nn
\eeqa
Similar cancellation between  the reduction of boundary action and the total derivative terms in the bulk action has been observed for the couplings at order $\alpha'$ in \cite{Garousi:2021cfc}.

Using the generalized metric $\cS$
\beqa
\cS\equiv \eta \left(\matrix{G^{-1}& -G^{-1}B&\cr BG^{-1}&G-BG^{-1}B&}\right)\labell{S}
\eeqa
where $\eta$ is the  metric of the $O(d,d)$ group which in the non-diagonal form is 
\beqa
\eta&=& \left(\matrix{0& 1&\cr 1&0&}\right),
\eeqa
one can write the bulk action as
\beqa
\bS_0^c&=&-\frac{2}{\kappa^2}\int dt e^{-\Phi}\Bigg[-\frac{1}{8}\tr(\dS^2)-\dP^2\Bigg]
\eeqa
which is invariant under the global $O(d,d,R)$ transformations because the one-dimensional dilaton is invariant and the generalized metric transforms as
\beqa
\cS&\rightarrow &\Omega^T\cS\Omega
\eeqa
where $\Omega$ belong to the  $O(d,d,R)$ group, \ie $\Omega^T\eta\Omega=\eta$. Note that $\tr(\dS)=0$. Hence the reduction of the  extrinsic curvature \reef{cK} can not be written in $O(9,9)$ invariant form. So it was nessary that this term was cancelled with the total derivative term in the bulk action. In other words, there is no way to write the boundary action in $O(d,d)$ invariant form unless it is zero.
 
\section{Cosmological reduction at order $\alpha'^3$}

 The NS-NS couplings in the bulk effective action of type II superstring theory at order $\alpha'^3$ have been found in \cite{Garousi:2020gio,Garousi:2020lof} by imposing the $Z_2$-symmetry on the minimal gauge invariant couplings. In the particular scheme used in \cite{Garousi:2020lof}, the dilaton appears only as the overall factor $e^{-2\phi}$, and metric and $B$-field appear in the Riemann curvature, $H$ and the first covariant derivative of $H$, \ie
 \beqa
 \quad \bS_3= -\frac{2c}{\kappa^2}\int d^{10} x\sqrt{-g} \Big[e^{-2\phi} L_3(G,B)+ \cdots\Big]\labell{S3bf}
\eeqa
where dots represent the R-R and fermion fields in which we are not interested, and $c$ is an overall factor which can not be fixed by the T-duality constraint.  The gravitational sector is 
\beqa
L_3(G)&=& 2 R_{\alpha}{}^{\epsilon}{}_{\gamma}{}^{\varepsilon} R^{\alpha \beta \gamma \delta} R_{\beta}{}^{\mu}{}_{\epsilon}{}^{\zeta} R_{\delta \zeta \varepsilon \mu} + R_{\alpha \beta}{}^{\epsilon \varepsilon} R^{\alpha \beta \gamma \delta} R_{\gamma}{}^{\mu}{}_{\epsilon}{}^{\zeta} R_{\delta \zeta \varepsilon \mu}\labell{RRf}
\eeqa  
The couplings  in this sector are exactly the couplings that have been found by the S-matrix and sigma-model calculations \cite{Gross:1986iv,Gross:1986mw,Grisaru:1986vi,Freeman:1986zh} provided that   one chooses the overall parameter to be $c=-\z(3)/2^6$. There are 249  couplings  which involve $H$. They appear in 8 structures. There are two couplings with structure $H^8$, \ie
\beqa
L_3^{H^8}&=&\frac{1}{48} H_{\alpha}{}^{\delta \epsilon} H^{\alpha \beta 
\gamma} H_{\beta}{}^{\varepsilon \mu} H_{\gamma}{}^{\zeta 
\eta} H_{\delta \varepsilon}{}^{\theta} H_{\epsilon 
\zeta}{}^{\iota} H_{\theta \iota \kappa} H_{\mu 
\eta}{}^{\kappa}\nn\\&& -  \frac{9}{128} H_{\alpha}{}^{\delta 
\epsilon} H^{\alpha \beta \gamma} H_{\beta}{}^{\varepsilon 
\mu} H_{\gamma}{}^{\zeta \eta} H_{\delta 
\varepsilon}{}^{\theta} H_{\epsilon \zeta}{}^{\iota} H_{\eta 
\theta \kappa} H_{\mu \iota}{}^{\kappa}\labell{H8}
\eeqa 
There is one coupling with structure $RH^6$, \ie
\beqa
L_3^{RH^6}&=&\frac{9}{8} H_{\alpha}{}^{\delta \epsilon} H^{\alpha \beta \gamma} 
H_{\beta}{}^{\varepsilon \mu} H_{\gamma}{}^{\zeta \eta} H_{\delta 
\varepsilon}{}^{\theta} H_{\epsilon \zeta}{}^{\iota} R_{\mu \iota \eta \theta}\labell{RH6}
\eeqa
There are 7 couplings with structure $R^2H^4$, \ie
\beqa
L_3^{R^2H^4}&=&\frac{7}{2} H_{\alpha}{}^{\delta \epsilon} H^{\alpha \beta 
\gamma} H_{\beta}{}^{\varepsilon \mu} H_{\delta}{}^{\zeta 
\eta} R_{\gamma}{}^{\theta}{}_{\varepsilon 
\zeta} R_{\epsilon \theta \mu \eta}+\cdots
\eeqa
where dots refer to the other 6 couplings in this structure.  There are 22 couplings with structure $R^3H^2$, \ie
\beqa
L_3^{R^3H^2}&=&- \frac{15}{2} H^{\alpha \beta \gamma} H^{\delta \epsilon \
\varepsilon} R_{\alpha \delta}{}^{\mu \zeta} R_{\beta \mu 
\epsilon}{}^{\eta} R_{\gamma \eta \varepsilon \zeta}+\cdots
\eeqa
There are 77 couplings with the structure $(\nabla H)^2H^4$, \ie
\beqa
L_3^{(\prt H)^2H^4}&=&\frac{5}{8} H_{\alpha}{}^{\delta \epsilon} H^{\alpha \beta \gamma} H_{\beta}{}^{\varepsilon \mu} H_{\delta \varepsilon}{}^{\zeta} \nabla_{\epsilon}H_{\gamma}{}^{\eta \theta} \nabla_{\zeta}H_{\mu \eta \theta}+\cdots
\eeqa
There are 106 couplings with  the structure $R(\nabla H)^2H^2$, \ie
\beqa
L_3^{R(\prt H)^2 H^2}&=&\frac{457}{48} H_{\alpha}{}^{\delta \epsilon} H^{\alpha \beta 
\gamma} R_{\varepsilon \zeta \mu \eta} 
\nabla_{\delta}H_{\beta}{}^{\varepsilon \mu} 
\nabla_{\epsilon}H_{\gamma}{}^{\zeta \eta}+\cdots
\eeqa
There are 22 couplings with the structure $R^2(\nabla H)^2$, \ie
\beqa
L_3^{R^2(\prt H)^2}&=&- \frac{5}{24} R_{\epsilon \mu \varepsilon \zeta} R^{\epsilon 
\varepsilon \mu \zeta} \nabla_{\delta}H_{\alpha \beta \gamma} 
\nabla^{\delta}H^{\alpha \beta \gamma} +\cdots
\eeqa
And finally, there are 12 couplings with the structure $(\nabla H)^4$: 
\beqa
L_3^{(\prt H)^4}&=&\frac{1}{8} \nabla^{\delta}H^{\alpha \beta \gamma} 
\nabla_{\epsilon}H_{\gamma}{}^{\mu \zeta} 
\nabla_{\varepsilon}H_{\delta \mu \zeta} 
\nabla^{\varepsilon}H_{\alpha \beta}{}^{\epsilon}+\cdots
\eeqa
We refer the interested readers to \cite{Garousi:2020lof} for the explicit form of all couplings. 

To find the cosmological reduction of these couplings we first find the cosmological reduction of the Riemann curvature, $H$ and $\nabla H$. They are
\beqa
&&R_{ijkl}=-\frac{1}{4}\dG_{il}\dG_{jk}+\frac{1}{4}\dG_{ik}\dG_{jl}\,\,;\,\, R_{i0jk}=0\,\,;\,\,R_{i0j0}=\frac{1}{4}\dG_{ik}\dG^k{}_j-\frac{1}{2}\ddot{G}_{ij}\nn\\
&&H_{ijk}=0\,\,;\,\, H_{ij0}=\dB_{ij}\,\,;\,\, \nabla_0H_{ijk}=0\,\,;\,\, \nabla_k H_{ij0}=0\\
&&\nabla_l H_{ijk}=-\frac{1}{2}\dB_{jk}\dG_{il}+\frac{1}{2}\dB_{ik}\dG_{jl}-\frac{1}{2}\dB_{ij}\dG_{kl}\,\,;\,\,\nabla_0H_{ij0}=-\frac{1}{2}\dB_j{}^k\dB_{ik}-\frac{1}{2}\dB_i{}^k\dG_{jk}+\ddot{B}_{ij}\nn
\eeqa
Using the above reductions,  one finds the following reduction for the Lagrangians $L_3^{H^8}$:
\beqa
L_3^{H^8}&=&13 (\Tr(M^4))^2/64 + 61 \Tr(M^8)/128
\eeqa
where the $9\times 9$ matrices $M=G^{-1}\dB$ and $L=G^{-1}\dG$. The reduction of the Lagrangian $L_3(G)$ is the following
\beqa
L_3(G)&=&(\Tr(L^4))^2/64 + 5\Tr(L^8)/128+\cdots
\eeqa
where dots represent terms which have $\ddot{G}$, $\Tr(L^2)$, $\Tr(L)$ or $\Tr(L^3)$. The reduction of all other Lagrangians are
\beqa
L_3^{RH^6}&=&9 \Tr(L^2M^6)/16 + 9 \Tr(LM^2LM^4)/16 - 9 \Tr(LM^3LM^3)/
   32\nn\\&& + 9 \Tr(L^2M^2) \Tr(M^4)/16+\cdots\nn\\
L_3^{R^2H^4}&=&  19 \Tr(L^2M^2)^2/64 + 45 \Tr(L^2M^2L^2M^2)/64 - 15 \Tr(L^2M^2LMLM)/
   16\nn\\&& + 17 \Tr(L^2M^3L^2M)/32 + 7 \Tr(L^2MLM^2LM)/16 - 5 \Tr(L^3M^2LM^2)/
   32 \nn\\&&- 7 \Tr(L^3M^3LM)/16 + 17 \Tr(L^4M^4)/32 - 11 \Tr(L^2M^2) \Tr(LMLM)/
   32 \nn\\&&+ \Tr(L^4) \Tr(M^4)/128+\cdots\nn\\
L_3^{R^3H^2}&=& -81 \Tr(L^3ML^3M)/64 - 89 \Tr(L^2M^2) \Tr(L^4)/128 + 147 \Tr(L^4ML^2M)/
  64\nn\\&& - 99 \Tr(L^5MLM)/64 + 23 \Tr(L^6M^2)/32 + 75 \Tr(L^4) \Tr(LMLM)/256 +\cdots \nn\\
 L_3^{(\prt H)^2H^4}&=& 2257 \Tr(L^2M^6)/192 - 509 \Tr(LM^2LM^4)/64  + 91 \Tr(LMLM) \Tr(M^4)/96\nn\\&& + 169 \Tr(LM^5LM)/192 + 33 \Tr(L^2M^2) \Tr(M^4)/
  64- 389 \Tr(LM^3LM^3)/
  192+\cdots\nn\\
 L_3^{R(\prt H)^2 H^2}&=& 3 (\Tr(L^2M^2))^2/8 - 23 \Tr(LMLMLMLM)/48 + 1969 \Tr(L^2M^2LMLM)/
  192\nn\\&& + 933 \Tr(L^2M^3L^2M)/64 - 473 \Tr(L^2MLM^2LM)/
  96 - 1757 \Tr(L^3M^2LM^2)/192\nn\\&& - 17 \Tr(L^3M^3LM)/2 + 1391 \Tr(L^4M^4)/
  192 - 1385 \Tr(L^2M^2) \Tr(LMLM)/384\nn\\&& - 229 (\Tr(LMLM))^2/
  96 + 383 \Tr(L^2M^2L^2M^2)/96 - 311 \Tr(L^4) \Tr(M^4)/128+\cdots\nn\\
L_3^{R^2(\prt H)^2}&=& 17 \Tr(L^3ML^3M)/16 - \Tr(L^2M^2) \Tr(L^4)/16 + 29 \Tr(L^4ML^2M)/
  16\nn\\&& + 57 \Tr(L^5MLM)/64 + 3 \Tr(L^6M^2)/4 + 27 \Tr(L^4) \Tr(LMLM)/256 +\cdots\nn\\
 L_3^{(\prt H)^4}&=& 3 (\Tr(L^2M^2))^2/64 + 3 \Tr(L^2M^2L^2M^2)/64 - 113 \Tr(LMLMLMLM)/192 \nn\\&&+ 163 \Tr(L^2M^3L^2M)/48 - 865 \Tr(L^2MLM^2LM)/
  96 - 1247 \Tr(L^3M^2LM^2)/192 \nn\\&&- 41 \Tr(L^3M^3LM)/64 + 2383 \Tr(L^4M^4)/
  192 + 1439 \Tr(L^2M^2) \Tr(LMLM)/384\nn\\&& + 113 (\Tr(LMLM))^2/
  96 - 2485 \Tr(L^2M^2LMLM)/
  192\nn\\&& - 277 \Tr(L^4) \Tr(M^4)/128+\cdots
   \eeqa
 where dots represent terms which have $\ddot{G}$, $\ddot{B}$, $\Tr(L^2)$, $\Tr(M^2)$, $\Tr(L)$ or $\Tr(L^3)$.  As it has been argued in \cite{Hohm:2015doa}, using  the field redefinitions and total derivative terms, the couplings involving these structures can be converted to the other couplings which have no such structures. Before showing how this works, let us add  all above reductions to  find the cosmological reduction of the Lagrangian $L_3(G,B)$, \ie
 \beqa
 L_3(G,B)&=&23 (\Tr(L^2M^2))^2/32 + 455 \Tr(L^2M^2L^2M^2)/96 - 29 \Tr(L^2M^2LMLM)/
  8 \nn\\&&+ 3553 \Tr(L^2M^3L^2M)/192 + 2365 \Tr(L^2M^6)/192 - 27 \Tr(L^2MLM^2LM)/
  2 \nn\\&& - 1517 \Tr(L^3M^2LM^2)/96 - 613 \Tr(L^3M^3LM)/64 - 13 \Tr(L^3ML^3M)/
  64 \nn\\&& - 97 \Tr(L^2M^2) \Tr(L^4)/128 + 
 (\Tr(L^4))^2/64 + 323 \Tr(L^4M^4)/16  \nn\\&&+ 263 \Tr(L^4ML^2M)/64 - 21 \Tr(L^5MLM)/
  32 + 47 \Tr(L^6M^2)/32 + 5 \Tr(L^8)/128 \nn\\&& - 473 \Tr(LM^2LM^4)/
  64 - 443 \Tr(LM^3LM^3)/192 + 169 \Tr(LM^5LM)/
  192 \nn\\&& - 13 \Tr(L^2M^2) \Tr(LMLM)/64 + 51 \Tr(L^4) \Tr(LMLM)/
  128 - 29 (\Tr(LMLM))^2/24  \nn\\&&- 205 \Tr(LMLMLMLM)/
  192 + 69 \Tr(L^2M^2) \Tr(M^4)/64 - 587 \Tr(L^4) \Tr(M^4)/
  128  \nn\\&&+ 91 \Tr(LMLM) \Tr(M^4)/96 + 13 (\Tr(M^4))^2/64 + 61 \Tr(M^8)/128+\cdots\labell{cL3}
 \eeqa
 which has 28 structures which have no  $\ddot{G}$, $\ddot{B}$, $\Tr(L^2)$, $\Tr(M^2)$, $\Tr(L)$ or $\Tr(L^3)$. The cosmological reduction of the effective action \reef{S3bf}, then is
 \beqa
 \bS_3^c&=&-\frac{2c}{\kappa^2}\int dt \Bigg[e^{-\Phi}L_3(G,B)+\cdots\Bigg]\labell{cS3}
 \eeqa
where $L_3(G,B)$ is given in \reef{cL3}.

To remove the total derivatives terms and field redefinition freedom from \reef{cS3}, we  add all total derivative terms at order $\alpha'^3$ and all field redefinitions with arbitrary coefficients to  \reef{cS3}. We add the following total derivative terms:
\beqa
-\frac{2c}{\kappa^2}\int dt e^{-\Phi}\cJ_3&\equiv&-\frac{2c}{\kappa^2}\int dt\frac{d}{dt}(e^{-\Phi}\cI_3)
\eeqa
where  $\cI_3$ is   all possible    terms at seven-derivative level with even parity which are constructed from $\dP$, $\dB$, $\dG$, $\ddot{\Phi}$, $\ddot{B}$, $\ddot{G}$,  $\cdots$.  Using the package   "xAct" \cite{Nutma:2013zea}, one finds there are 2288 such terms, \ie
\beqa
\cI_3&=&j_1\dB_i{}^j\dB_j{}^k\dB_k{}^l\dG_l{}^m\dB_m{}^n\dB_n{}^p\dB_p{}^i+\cdots
\eeqa
where the coefficients  $J_1,\cdots, J_{2288}$ are 2288 arbitrary parameters.

One can change the field variables in \reef{creduce} as 
\begin{eqnarray}
G_{ij}&\rightarrow &G_{ij}+\alpha'^3 \delta G^{(3)}_{ij}\nn\\
B_{ij}&\rightarrow &B_{ij}+ \alpha'^3\delta B^{(3)}_{ij}\nn\\
\Phi &\rightarrow &\Phi+ \alpha'^3\delta\Phi^{(3)}\nn\\
n &\rightarrow &n+ \alpha'^3 \delta n^{(3)}\labell{gbpn}
\end{eqnarray}
where the matrices  $\delta G^{(3)}_{ij}$, $\delta B^{(3)}_{ij}$ and $\delta\Phi^{(3)}, \delta n^{(3)}$ are all possible  terms at 6-derivative level constructed from $\dP$, $\dB$, $\dG$, $\ddot{\Phi}$, $\ddot{B}$, $\ddot{G}$,  $\cdots$.  The perturbations  $\delta G^{(3)}_{ij}$, $\delta\Phi^{(3)}$, $\delta n^{(3)}$ contain even-parity terms and $\delta B^{(3)}_{ij}$ contains odd-parity terms, \ie
\beqa
\delta n^{(3)}&=&n_1\dB_i{}^j\dB_j{}^k\dB_k{}^l\dB_l{}^m\dB_m{}^n\dB_n{}^i+\cdots\nn\\
\delta \Phi^{(3)}&=&e_1\dB_i{}^j\dB_j{}^k\dB_k{}^l\dB_l{}^m\dB_m{}^n\dB_n{}^i+\cdots\nn\\
\delta G^{(3)}_{ij}&=&d_1\dB_i{}^k\dB_k{}^l\dB_l{}^m\dB_m{}^n\dB_n{}^p\dB_p{}_j+\cdots\nn\\
\delta B^{(3)}_{ij}&=&f_1\dG_i{}^k\dB_k{}_j\dB_l{}^m\dB_m{}^n\dB_n{}^p\dB_p{}_l+\cdots
\eeqa
The coefficients $n_1,\cdots, n_{748}$, $e_1,\cdots, e_{748}$, $d_1,\cdots, d_{1105}$  and $f_1,\cdots, f_{665}$ are arbitrary parameters. When the field variables in $\!\!\bS_3^c$  are changed according to the above  field redefinitions, they produce some couplings at orders $\alpha'^6$ and higher in which we are not interested in this paper. However, when the NS-NS field variables in $\!\!\bS_0^c$  are changed,  the following   couplings  at order $\alpha'^3$ are produced:
\beqa
\delta\!\!\bS_0^c&=&-\frac{2\alpha'^3}{\kappa^2}\int dt e^{-\Phi}\Bigg[\delta n^{(3)}\left(-\frac{1}{4}\dB_{ij}\dB^{ij}-\frac{1}{4}\dG_{ij}\dG^{ij}+\dP^2\right)\labell{dS0c}\\
&&+\delta \Phi^{(3)}\left(-\frac{1}{4}\dB_{ij}\dB^{ij}-\frac{1}{4}\dG_{ij}\dG^{ij}+\dP^2\right)-2\dP\frac{d}{dt}\delta \Phi^{(3)}\nn\\
&&+\delta G^{(3)}_{ij}\left(-\frac{1}{2}\dB_k{}^j\dB^{ki}-\frac{1}{2}\dG_k{}^j\dG^{ki}\right)+\frac{1}{2}\dG^{ij}\frac{d}{dt}\delta G^{(3)}_{ij}+\frac{1}{2}\dB^{ij}\frac{d}{dt}\delta B^{(3)}_{ij}\Bigg]\nn\\
&\equiv& -\frac{2\alpha'^3c}{\kappa^2}\int dt e^{-\Phi}{\cal K}_3\nn
\eeqa
where we have used the fact that the lapse function appears in the action \reef{S0c} by replacing $dt\rightarrow dt/n$ \cite{Hohm:2015doa}.

Adding the total derivative terms and the field redefinition terms  to the action \reef{cS3}, one finds new  action $S_3^c$, \ie
 \beqa
 S_3^c= -\frac{2 c}{\kappa^2}\int dt \Bigg[ e^{-\Phi}\cL_3(G,B,\Phi)+\cdots\Bigg]\labell{S3bf1}
\eeqa
where the  Lagrangian  $\cL_3(G,B,\Phi)$ is related to the Lagrangian $\cL_3(G,B)$ as   
\beqa
\cL_3&=&L_3+{\cal J}_3+{\cal K}_3\labell{DLK}
\eeqa
The action $\bS_3^c$ and $S_3^c$ are physically equivalent. They appear in different schemes.  Choosing different values for the arbitrary parameters in ${\cal J}_3$, ${\cal K}_3$, one would find different forms of couplings for the Lagrangian $\cL_3$.  We choose these parameters such that all terms that have any derivative of  $\Phi$, second  and higher derivatives of $G,B$, and  all terms that have $\Tr(L^2)$, $\Tr(M^2)$, $\Tr(L)$ or $\Tr(L^3)$ to be zero. Inserting  the resulting relations between the parameters into \reef{DLK}, we find the following scheme for the Lagrangian $\cL_3$:
\beqa
\cL_3&=& (\Tr(L^2M^2))^2/4 - 3 \Tr(L^2M^2L^2M^2)/32 + 3 \Tr(L^2M^2LMLM)/
    8 - 3 \Tr(L^2M^3L^2M)/16\nn\\&& + 3 \Tr(L^2M^6)/16 - 3 \Tr(L^2MLM^2LM)/
    16 - 3 \Tr(L^3M^2LM^2)/16 + 3 \Tr(L^3M^3LM)/8\nn\\&& - 3 \Tr(L^3ML^3M)/
    32 - \Tr(L^2M^2) \Tr(L^4)/8 + 
   (\Tr(L^4))^2/64 - 3 \Tr(L^4M^4)/16\nn\\&& + 3 \Tr(L^4ML^2M)/16 - 3 \Tr(L^5MLM)/
    16 + 3 \Tr(L^6M^2)/16 - 3 \Tr(L^8)/128\nn\\&& + 3 \Tr(LM^2LM^4)/
    16 - 3 \Tr(LM^3LM^3)/32 - 3 \Tr(LM^5LM)/16 \nn\\&&- \Tr(L^2M^2) \Tr(LMLM)/
    4 + \Tr(L^4) \Tr(LMLM)/16 + 
   (\Tr(LMLM))^2/16\nn\\&& - 3 \Tr(LMLMLMLM)/64 - \Tr(L^2M^2) \Tr(M^4)/
    8 + \Tr(L^4) \Tr(M^4)/32 \nn\\&&+ \Tr(LMLM) \Tr(M^4)/16 + 
   (\Tr(M^4))^2/64 - 3 \Tr(M^8)/128\labell{cL}
\eeqa
which has the same 28 structures as in \reef{cL3} but with different coefficients. Note that in this scheme, there is no  term in $\cL_3$ other than the above 28 couplings.

Now using the definition of the generalized metric in \reef{S}, one finds
\beqa
\tr(\dS^4)&=&2 \Tr(L^4) + 2 \Tr(M^4) - 8 \Tr(L^2M^2) + 4 \Tr(LMLM)\nn\\
\tr(\dS^8)&=&8 \Tr(L^2M^2L^2M^2) - 32 \Tr(L^2M^2LMLM) + 16 \Tr(L^2M^3L^2M) - 16 \Tr(L^2M^6)\nn\\&& + 
 16 \Tr(L^2MLM^2LM) + 16 \Tr(L^3M^2LM^2) - 32 \Tr(L^3M^3LM) + 8 \Tr(L^3ML^3M)\nn\\&& + 
 16 \Tr(L^4M^4) - 16 \Tr(L^4ML^2M) + 16 \Tr(L^5MLM) - 16 \Tr(L^6M^2) + 2 \Tr(L^8)\nn\\&& - 
 16 \Tr(LM^2LM^4) + 8 \Tr(LM^3LM^3) + 16 \Tr(LM^5LM)\nn\\&& + 4 \Tr(LMLMLMLM) + 2 \Tr(M^8)
\eeqa
Using the above $O(9,9)$-invariant expressions, one can write \reef{cL} as 
\beqa
\cL_3&=&\frac{1}{256}(\tr(\dS^4))^2-\frac{3}{256}\tr(\dS^8)\labell{cL3f}
\eeqa
which is consistent with the cosmological action \reef{cosm}. We have done the same calculations with couplings in \cite{Garousi:2020gio}, and found exactly the same result. The form of the one-dimensional field redefinitions and total derivative terms, however, are different in the two cases. The above calculations confirm the NS-NS couplings at order $\alpha'^3$ which has been found in \cite{Garousi:2020gio,Garousi:2020lof}.



\end{document}